**Title.** Anticipation in architectural experience: a computational neurophenomenology for architecture?

**Authors:**

Zakaria Djebbara[1]* (0000-0003-2370-6092)

Thomas Parr[2] (0000-0001-5108-5743)

Karl Friston[2] (0000-0001-7984-8909)

[1] Department of Architecture, Design and Media Technology, Aalborg University, Aalborg, Denmark

[2] Wellcome Centre for Human Neuroimaging, University College London, London, UK

*Corresponding author: zadj@create.aau.dk

**Abstract.** The perceptual experience of architecture is enacted by the sensory and motor system. When we act, we change the perceived environment according to a set of expectations that depend on our body and the built environment. The continuous process of collecting sensory information is thus based on bodily affordances. Affordances characterize the fit between the physical structure of the body and capacities for movement in the built environment. Since little has been done regarding the role of architectural design in the emergence of perceptual experience on a neuronal level, this paper offers a first step towards the role of architectural design in perceptual experience. An approach to synthesize concepts from computational neuroscience with architectural phenomenology into a computational neurophenomenology is considered. The outcome is a framework under which studies of architecture and cognitive neuroscience can be cast. In this paper, it is first argued that the experience of space is an embodied process—realized through action-perception as directed by affordances. Second, we integrate a sensorimotor contingency theory with a predictive coding architecture of the brain that in turn links the perceptual experience of forms and action possibilities with neuronal processes. Here, we argue that the sum of action possibilities and the inferred precision thereof can reflect the understanding of the designed space, while at the same time underwrite the basis for the perceptual experience. To this end, affordances are inherently related to perceptual experience. Finally, by reviewing recent empirical evidence we propose a principle of anticipation in architectural experience.

**Keywords.** Architectural experience, Sensorimotor dynamics, Active inference, Mobile Brain/Body Imaging, Affordances, Process philosophy

**Declarations.** The authors declare no conflicts of interest.

**Acknowledgments.** We would like to thank Andrea Jelic, Tenna Doktor Olsen Tvedebrink, Runa Hellwig, and Lars Brorson Fich from the Brain, Body, and Architecture (BBAR) group at Aalborg University, Denmark for important discussions and insights. We would further like to thank Professor Gramann and his team at Berlin Mobile Brain/Body Imaging Lab (BeMoBIL) at Technische Universität Berlin.

## 1. Introduction

Concretized in space through materials, architecture can synchronously awaken all the elements and complexities of perception. The continuous experience of the textured stone surface, light changing with movement, resonant sounds of space, and the bodily relation of scale and proportion are how the built environment speaks through the silent process of perceptual phenomena (Holl, Pallasmaa and Pérez-Gómez, 2006). This continuous process of architectural experience is understood as a 'whole' (Pallasmaa, 2011, 2012), i.e. despite our ability to disassemble the many sensory elements within such a process, the phenomenological account states they all merge into a collection of sensations manifested as an enmeshed experience. The 'whole' reflects the enmeshment of sensations. Thus, as one propels through space, the body, entirely, becomes the measure of the architectural experience. Henceforth, the architectural experience is defined as follows:

**Architectural experience**: A 'whole', enmeshed experience of space that emerges from the process of continuous overlapping sensations.

During an architectural experience, the human body and brain can dissect the 'whole' and permeate sensations to become consciously aware of the perception as organized fragments. Indeed, the architectural experience is constituted by a series of perceptual experiences that can be analyzed consciously by intention (see for instance Husserl, 1997)—yet, the continuous stream of sensation endures unconsciously. Nevertheless, conscious or unconscious, as elaborated in this paper, the reading of our environment through architectural experience rests on a biological purpose. This illuminates a major challenge facing phenomenology. It concerns the limitations of theoretical argumentations in the advancement of the underlying biological functions for such phenomenal experiences (Noë, 2007). Phenomenology runs short of persuasive argumentation by remaining in a purely theoretical territory. Thus, a biological sensorimotor account of perception is sought that speaks to the traditional phenomenological accounts of experience.

There are at least two complementary advantages of having a practical approach to perception as provided by a sensorimotor contingency (SMC) account (O'Regan and Noë, 2001a). First, by building on the brain-body-environment interactions (Varela, Thompson and Rosch, 2016), SMC claims that perception requires no internal representation[1], but rather a practical knowledge of how to realize the predicted perception. Appropriately, SMC has been portrayed as a radical type of enactivism (Hutto and Myin, 2013). The central argument in SMC draws on practical 'knowing-how' knowledge over propositional 'knowing-that' knowledge (Ryle, 1945). For instance, the incomplete perception of the built environment—despite perspectival constraints—is not based on an internal representation of the space. Instead, the built environment is experienced as complete because we know how to move our body and eyes to reveal the predictable sensory information about the built environment. Second, according to SMC, practical knowledge is to exercise mastery of sensorimotor laws. Each modality varies with action in each their way—in this sense, the dynamics of visual sensation combined with motor execution provides a foundation for the perceptual experience. Perceptual experience, infamously formulated as 'quale', can now be said to emerge through action where the qualitative features are aspects thereof. It is precisely due to the exercising of skillfull mastery of sensorimotor laws providing sensory predictability that shapes the basis for perception. Perceptual experience in this sense is rejected as a state or static property while instead put forward as a process of change if action unfolds so and so—hence, sensorimotor contingency.

Although perceptual experience is typically examined by the experience of colors, in this paper, we insist that the 'form' in space holds a quale status. Notably, form is an essential component in the architectural design and essentially integrated with how affordances emerge (Rietveld and Kiverstein, 2014; Pezzulo and Cisek, 2016; Djebbara *et al.*, 2019). The form of space speaks to all sensory systems and thereby reveals how one may explore the space with the whole body (Djebbara, Fich and Gramann, 2020). Along the same lines that SMC argues the perceptual experience to depend on the sensorimotor patterns that the eyes may reveal, the perceptual experience of form is a matter of how the whole body moves through space, hauling all sensory systems along, exploring the qualities that they all may reveal. This is indeed in line with phenomenology—however, what phenomenology could not supply, in terms of the level of detail concerning the emergence of perceptual experience, is instead supplied by SMC. Nevertheless, the heavy-lifting of SMC rests on how sensorimotor dynamics operationalize, which remains obscure in the influential paper by O'Regan and Noë (2001a; Buhrmann, Di Paolo and Barandiaran, 2013). This particular challenge is proposed to be resolved (in part) by coupling SMC with 'active inference', which is a biophysical and computational approach to action and perception. Contingencies of sensorimotor activity beyond 'form' become thus active inferences, or simply predictions of a certain kind; i.e., "What would happen if I did that?"

---

[1] The discussion on whether the brain makes use of representations is beyond the scope of this paper.

Conceiving experience as a process is conveniently known in philosophy as 'process philosophy'[2]. At the center of process philosophy, continuity and change characterize the nature of human experience (Rescher, 2000; Bergson, 2001). The continuity stems from human perception being governed by a constant influx of sensory information, which is simultaneously responsible for the constant change. Indeed, the rate of change can be conceived as the dynamics of perception—however, the change in perception does not hold a single source. In computational neuroscience, by conceiving the functions of the brain as Bayesian belief updating (a.k.a., evidence accumulation), the process of perception may be initiated by either action under predicted (inferred) sensory signals based on prior experience (Friston et al., 2017). The challenging inference problem—faced by such generative models—is to determine which of the many available actions in the environment best change the sensory signals according to the expected perceptual consequences. The available actions are valued by the affordances of the environment (Gibson, 1977). Affordances refer to the possibilities for use, intervention, and action which the physical world offers and are determined by the fit between a body's structure, skills, and capacities for movement (Clark, 1999). Importantly, since one action may be afforded over another, it is worth noting that affordances differ in degree, while action possibilities differ in kind (Friston et al., 2012). The nature of affordances in active inference is slightly more nuanced than Gibsonian affordances in the following sense: Gibsonian affordance is an attribute of states of affairs in the world 'out there' relative to the way one engages with the world. In active inference, affordance becomes an attribute of action possibilities that have to be inferred. Crucially, these affordances can be pragmatic (e.g., "sitting on this is the kind of thing I do") or epistemic (e.g., "I should go upstairs to find somewhere to sit").

This paper proposes to combine active inference with the SMC theory of perceptual experience to illuminate how architectural experience arises. From the combination, we propose that affordances attune sensory signals that in turn form the basis for the perceptual experience of form in designed spaces. The argument is straightforward:

> **The premise of phenomenology.** Architectural experience is based on perceptual experience.
>
> **The premise of sensorimotor contingency theory.** The perceptual experience of form in space depends on sensorimotor laws.
>
> **The premise of active inference.** Sensorimotor laws are shaped by (i) the possible sensations elicited by actions and (ii) the predicted sensations led by prior experiences.

---

[2] Other process philosophers include Gottfried Leibniz, George Herbert Mead, Henri Bergson, William James, John Dewey, Alfred North Whitehead and Charles Sanders Peirce.

**Conclusion**. Thus, the architectural experience is predicated on the anticipated actions afforded by designed forms in the environment.

Similar to how Gestalt principles hold for perceptual experiences, it is here argued that there exist architectural Gestalt principles that can be understood through the proposed enactive framework. Besides advancing knowledge of how the designed space interacts with the body and brain, the proposed approach demystifies how architectural experience can be investigated through methods other than descriptions. As it is proposed that the etiology of experience depends on sensory and motor information, a principle of anticipation emerges naturally from the combination of the theories. Possible actions, which essentially are anticipative, account for the perceptual experience and the grip on the environment. By establishing a computational approach to architectural experiences, it provides a promising approach to the quantification of affordances, which in turn attune perceptual experiences. Therefore, we suggest that a computational neurophenomenological approach, by quantifying the affordances, may inform designers about the possible experiences.

The aims of this paper are threefold. Through a philosophical discussion, the first aim entails coupling SMC to the experience of architecture, emphasizing how the experience of architectural form depends on the bodily structure and continuity in sensation and action. This task is taken up in Section 2. The second aim involves establishing how SMC engages with the predictive processing in the brain, depicted by (variational) active Bayesian inference. Such an outline is presented in Section 3 and considered within the discourse of cognitive neuroscience. The third and final aim is to clarify how the designed space affects perceptual experience through affordances, which are essentially predictions, and briefly review empirical experiments of architecture. Section 4 is dedicated to that final aim.

## 2. Phenomena of architectural experience

As one may wonder whether architects always have been aware of the effect caused by their design, indeed such wonder suggests a link to biological and evolutionary processes. Similar to Zeki's opinion on artists throughout history, one may wonder whether architects may have been "[…] neurologists, studying the brain with techniques that are unique to them and reaching interesting but unspecified conclusions about the organization of the brain. Or, rather, that they are *exploiting the characteristics of the parallel processing-perceptual systems of the brain to create their works*, sometimes even restricting themselves largely or wholly to one system, as in kinetic art" (Zeki, 1998, p. 77; emphasis added). With this in mind, the phenomenological tradition, within which architectural experience is typically embedded, provides a useful overview of how that might have been the case. As the phenomenology of architecture has been reviewed various times, this paper restricts itself to merely the relevant aspects to link SMC.

### 2.1. Embodiment

How is the architectural experience embodied? We answer this question by providing a brief phenomenological analysis of experience that in turn can be generalized. This is followed by the SMC account of perceptual experience.

Steen Eiler Rasmussen, a Danish architectural phenomenologist made a noteworthy impact on architectural theory with his book 'Experiencing Architecture' (2012/1957) regarding how the experience of architecture is entirely bodily. Rasmussen insisted on the importance of movement, perspective, and active perception for experiencing the built environment (2012, p. 35). To Rasmussen, the impressions of architectural quality are actively perceived in time and movement. Although the built environment itself enacts no rhythm as such, it is experienced depending on how it may afford fast or slow movement, for instance through curvy or perpendicular corners, steep or low ramps. The just-viewed scenes remain in retention while the upcoming scene is actively arranged through perceptual and motor predictions (Husserl, 2001a; Gallagher and Zahavi, 2012, chap.

4). With the series of perspectival deformations following this rhythm, perception develops according to an angle, velocity, and complexity of movement (Rasmussen, 2012, p. 137). Rasmussen essentially describes a process within which motor activity and sensation are continuous and constitutive of experience. This is an observation shared by other architectural phenomenologists. For instance, Holl, Pallasmaa, and Pérez-Gómez suggest that with movement, perspective, and active perception, as one opens a door "[…] our body weight meets the weight of the door; our legs measure the steps as we ascend the stairs, our hand strokes the handrail and our entire body moves diagonally and dramatically through space" (Holl, Pallasmaa and Pérez-Gómez, 2006, p. 35). Indeed, these elements as a 'whole' compose a complete enmeshed experience of space.

Architectural experience consists of a collection of perceptual experiences that are difficult to segregate from one another. Visual information is inherently intertwined with proprioceptive information, forming an enmeshed experience of the bodyweight meeting the door, all while the visual perspective continues to change as one moves forward. Instead of addressing all contributing elements, we restrict the proposal to 'form' in space. However, the short analysis above encircles the architectural experience that is sought to be linked with SMC and further with active inference. Several aspects of the above analysis are relevant to SMC. To make the congruence evident, an overview is provided of how phenomenology analyses (i) the sense of perceiving more than sensory information provides, and (ii) the entanglement of the body structure and motor activity with the process of moving from sensation to perception.

## 2.2. Phenomenology and SMC

Generally, phenomenology does not take for granted how the physical world enters into our experience (Merleau-Ponty, 1964, 1968, 2013; Husserl, 1997, 2001b; Bergson, 2004). Yet, phenomenological reflections can only bring one so far concerning the precondition of an object for perceptual experience. In fact, a major challenge facing phenomenology concerns the failure to demonstrate that one cannot have an experience of the same theoretical kind without the object:

"This claim about the world-involving character of perceptual experience – that experience is an encounter with things and situations – is not compatible with any old metaphysical or empirical picture of perception and its nature. For in presenting perceptual experience as a kind of involvement with or entanglement with situations and things, the phenomenology presents experience as something that could not occur in the absence of situations and things." (Noë, 2007)

To do justice to phenomenology, however, it goes further than mere reflections and descriptions of experiences. It does so by analyzing the entanglement of the body structure and motor activity with the process of sensation to perception, essentially addressing how the physical world enters into our experience (Merleau-Ponty, 1964, 1968, 2013; Husserl, 1997, 2001b; Bergson, 2004). Husserl, for instance, dedicated half the lectures given in 1907 (Husserl, 1997) to kinaesthetic and oculomotor changes in sensation during action, emphasizing the importance of movement to make sense of sensations. Notably, perception is not understood as the sum of sensation by phenomenologists. In truth, both Husserl and Merleau-Ponty observe the precondition of 'a point of view' for any perception. This is typically expressed by the phenomenological idiom "There are no view from nowhere", i.e. one cannot conceive of a place where oneself is not present in the situation providing a body to establish a point of view (Merleau-Ponty, 1964, p. 16). Therefore, phenomenology cast perception as situated and embodied by continuously using the body as an indexical 'here' concerning what appears in the environment (Zahavi and Overgaard, 2012, p. 285). Such an account essentially predicts that perception does not conform to the sum of sensations but is rather tuned by the situation—and the set of prior beliefs appropriate to that situation. This is indeed clear in Figure 1 where the elements in the center in both situations are identical, yet perceived as non-identical.

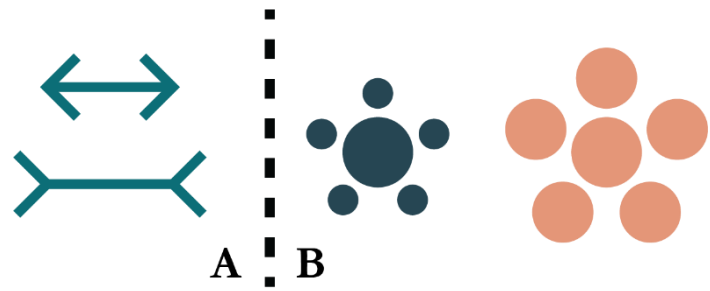

**Figure 1**. **A**. Müller-Lyer illusion. The central line is of equal width in both cases—yet, it is not experienced as such. **B**. Ebbinghaus-illusion. The two central circles in each group share the same size—yet, it is not experienced as such.

Suppose the elements in Figure 1 were physical structures, affording to be picked up and practically investigated. In such a situation, the structures are perceived differently from one another as they are picked up differently. If the central rods in Figure 1A were dissected from their arrows at the end, the rods would be similar in their physical structure, i.e. plain rods of equal length. They afford, then, to be practically investigated in the same manner (Gallagher and Zahavi, 2012, pp. 106–107). Perception is not divisible to simple sensory events or parts but must be grasped in its entirety within a setting from a view. Perception is immediate and pregnant with its form, devoid of deciphering, and intermediate inferences of what each sensory information provides (Merleau-Ponty, 1964, pp. 14–16). One perceives functionality beyond a single sensory modality—hence, the enmeshment.

To Husserl, the presence provides a totality open to a horizon of perspectives that merge and define the space in question. To this end, Husserl approached the constitution of perceptual experience through an investigation of the process of change in perception given a mobile body (Husserl, 1997, p. 133; Zahavi, 2003, p. 99). Notably, it was only during the investigation of temporality that Husserl realized the importance of movement in perception. This co-functionality is expressed by 'co-intentionality', which refers to the body continuously performing double functions to make sense of the perspectival changes (Husserl, 1997, pp. 131–132; Zahavi, 2003, pp. 99–100).

Parallel to the description of temporality (Figure 2), Husserl employs different terms, e.g. 'co-intentional', 'horizonal intentionality', and 'appresentation', to explain the emergence of (a 'whole') perception through the interaction of sensation and possible action (Husserl, 1999, p. 146). Essential to the concepts is the intentional presence, including the qualitative features, of spatial features only based on the anticipated possible actions related to the sensory modality. One thereby shows that the future (in part) depends on the action one chooses to unfold. Such sensorimotor dynamics are arguably described at a rather descriptive level by Husserl. In short, the nature of the co-intentionality integrates the situation and the body and thereby allows seeing more than sensory information provides, which in turn constitutes perception.

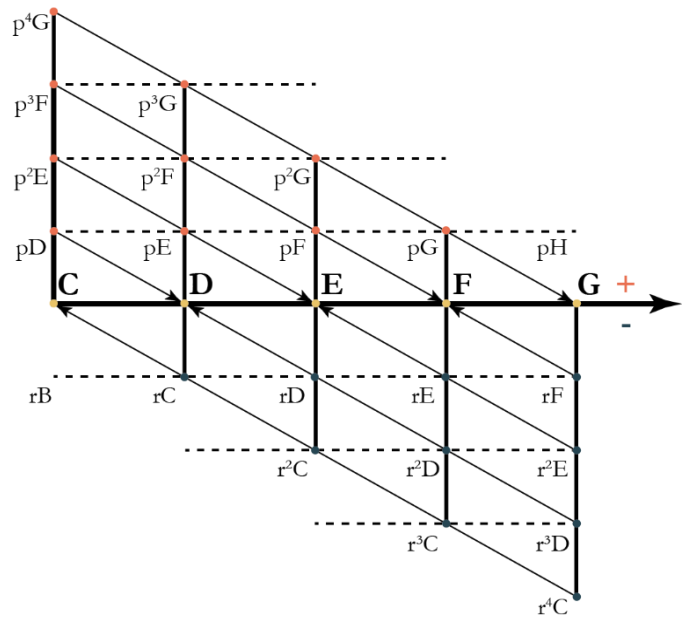

**Figure 2**. Husserl's diagram of time, based on drawings from the Bernauer Manuskripte (Husserl, 2001a, pp. 20–22). The solid horizontal line, CDEFG, designates the flow of events. Each phase contains retention (immediate past), a primal impression (present), and protention (immediate future). The retentions are designated by blue dots below the horizontal lines, whereas the protentions are red dots above the horizontal line. The vertical lines illustrate retention, protention, and primal impressions at the crossing. For instance, at the phase of D, there is a more certain protention of the upcoming phase, pE, compared to the further phases p²F and p³G. There is also retention of the prior phase, rC, which similarly is more certain than the further phases r²B and r³A.

Although phenomenology does not provide a detailed account of the sensorimotor coupling, there are numerous trails in the literature by Merleau-Ponty and Husserl. For instance, Merleau-Ponty observes that the multitude of perspectives of his apartment cannot be seen at the same time (2013, pp. 209–211). To the extent that one moves, the many faces of the apartment can only appear by the disappearance of another face of the apartment, and only throughout such process can one collect the necessary perceptual experiences to have a 'whole' conception of the apartment. In his description of Husserlian phenomenology, Zahavi provides an excellent example of the kinaesthetic horizon constitutes the 'whole' in the experience:

> "Whereas the actually given front of the wardrobe is correlated with a particular bodily position, the horizon of the cointended but momentarily absent profiles of the wardrobe (its backside, bottom, and so on) is correlated with my kinaesthetic horizon, that is, with my capacity for possible movement. The absent profiles are linked to an intentional if-then connection. If I move in this way, then this profile will become visually or tactually accessible. The absent backside of the wardrobe is only the backside of the same wardrobe I am currently perceiving because it can become present through a specific bodily movement." (Zahavi, 2003, p. 100)

The known actions that are not carried out are the actions that frame perception. These actions are sometimes referred to as Virtual actions (Bergson, 2001). Should one intend to invoke a specific perspective of the environment, then the Virtual actions that are latent within perception are summoned. Notably, intention can alter perception by action. In the tradition of phenomenology, the perceptual experience is discussed as 'intentionality' because the perceptual experience is to be aware of something in particular (Chalmers, 1996). Husserl (1997, 2001b) stresses the different kinds of intentionality correlated with the presence of things to constitute a type of perceptual experience. These involve signitive, imaginative, and perceptual intentions (Zahavi, 2003, p. 29). Such a contention resonates with SMC because it removes the use of representations when intending. The practical continuity of sensations and action in perception rests on the rejection of perceptual experience relying on internal representations. Instead, SMC suggests that the experience of the environment is based on the practical knowledge that the world is immediately accessible through exploration (O'Regan and Noë, 2001b). To this end, our perceptual organs are considered enacted (Thompson, 2007; Varela, Thompson and Rosch, 2016). According to SMC, the intended perception stems from the mastery of the change in sensory so that one perceives more than sensory information provides. However, SMC goes beyond and argues that the mastery of sensorimotor relations provides the ground on which the perceptual experience occurs.

## 2.3. The perceptual experience

Note that it is not the subjective experience that we zero in on, but rather the possibilities available for any perceptual experience. The mastery of sensorimotor changes addresses the possibility of perception in general—not to be confused with subjective experience. For instance, the subjective experience of one's childhood room is incomparable to that of another childhood room. The upshot of SMC is that perceptual experience can be generalized relative to perceptual organs. Indeed, this touches upon the technical term 'quale' in philosophy. Quale is typically described as the qualitative character of experience (Chalmers, 1996, pp. xi–xii). The term closely relates to the phenomenological discourse on intentionality (Gallagher and Zahavi, 2012, p. 123), i.e. to represent the world precisely like 'this', rather than 'that', as Nagel puts is: "[…] something that it is like to *be* that organism—something it is like *for* the organism" (1974, p. 436; original emphasis). The hard problem of qualia may be best described by Chalmers when questioning how physical structures as the body and brain can give rise to phenomenal experience (Chalmers, 1996, p. 5). A typical example to highlight the issue is to consider the perceptual experience of color.

Howbeit, within the framework of process theory, it is critical to understand the dynamics of the investigated subject. It is, therefore, necessary to examine the development of perceptual experience in time. Despite perceptual experience being typically examined through the experience of color, it is here suggested to examine the perceptual experience of form. To paraphrase Arnheim, it is the dynamics of architectural form that is discussed. Architectural form is revealed through diverse sensory organs. Tactility and echolocation by those robbed of sight and spatial shapes emerging due to differences in colors are examples that reveal form and rest on perceptual experiences. Forms are not necessarily restricted to the idea of lines but are rather a matter of dynamics in perspectives from situated points of view involving the 'whole' body. Indeed, form, arguably a relevant subject to the built environment, can be conceived as a relevant issue for quale.

Analyzing form as a function of time, Robbins manages to encapsulate the essence of process philosophy (Robbins, 2007, p. 24). Robbins introduces an experiment involving a rotating cube with varying speeds. When the speed is strobed in phase with its symmetry period, it is perceived as a rigid cube. However, when strobed out-of-phase and thereby breaking the temporal constraint of the perceptual system, the cube is perceived as wobbly (Robbins, 2007, pp. 5–6). The experiment is comparable to the Wagon-wheel effect. The dynamics of our perceptual system, which in this case refers to the visual system, are highly involved in the perceptual experience of form. With regards to architecture, spatial forms are perceived by the animate being through a dynamic figure-ground organization in perceptual groupings, e.g. color differences, lines, and general geometric forms. Thus, the claim is that the qualitative character is not a matter of static states of something, but directly an outcome of enacted perception. Experience, in this sense, is an active feature as put forward by O'Regan and Noë:

"Qualia are meant to be properties of experiential states or events. But experiences, we have argued, are not states. They are ways of acting. They are things we do. There is no introspectibly available property determining the character of one's experiential states, for there are no such states. Hence, there are, in this sense at least, no (visual) qualia. Qualia are an illusion, and the explanatory gap is no real gap at all […] Our claim, rather, is that it is confused to think of the qualitative character of experience in terms of the *occurrence* of something (whether in the mind or brain). Experience is something we do and its qualitative features are aspects of this activity." (O'Regan and Noë, 2001a, p. 960; original emphasis)

Since there are no states in experience but only dynamics, SMC suggests that the visual sensation is constituted by exercising our mastery of all kinds of sensorimotor contingencies making the issue of qualia obsolete (O'Regan and Noë, 2001a, p. 960; Hutto and Myin, 2013, p. 169). A circular space is sensed as circular because when acting, the dynamics of the form impact the visual system in such a way that it is recognized as circular. Form reveals a lot about the relation between observer and environment, for instance, through the perspectival view that holds allocentric and egocentric information. The skillful exercise consists of the practical knowledge in perception that is based on Virtual actions, i.e. if one were to act so and so, the sensory influx would vary so and so. In short, the sensation of a circular space is a way of doing things, and not something that emerges from the neuronal excitation in the visual cortex alone (O'Regan and Noë, 2001b, p. 83). It is important to emphasize that such contention of perceptual experience insists that the nature of the nervous influx of visual sensory is no different than that of auditory, olfactory, tactile, etc. Instead, what determines the sensory modality is the sensorimotor laws that govern the stimulation. Therefore, if tactile sensory influx were mimicking visual sensorimotor dynamics, the tactile sensation can provide a visual experience (O'Regan and Noë, 2001a, 2001b; Seth, 2014). In the very same sense that the experience of the color red depends on the affordances within perceptual systems, the experience of the space depends on the affordances of the perceptual system as a whole. Insofar, the examples are limited to visual perception, however, the grip of space and forms is certainly not limited to visual sensory, e.g. echolocation. The experience of the space depends thus on the whole body projecting itself through space, collecting, and enmeshing sensory information of all kinds to get a grip on the environment.

It is not that perception does not conform to sensations, but that the arrangement of the sensation is entirely embodied through motor capabilities. In turn, the collection of sensations does not constitute perception as a pure physical substratum. Accordingly, perceiving is " […] a condition whose nature depends *essentially* on the presence and involvement of the world encountered" (Noë, 2007; original emphasis). Point being, and in resonance with process philosophy, sensation and motor activity are not distinct processes followed by one another—instead, both are involved with specific situations and things. Sensation, through its texture that our gaze follows and joins with, is already the amplification of our motor being (Merleau-Ponty, 2013, p. 219). Indeed, the sensations that are constitutive of perceptual experiences must be understood with movement, while movement must be understood under possible behavior.

## 2.4. Process philosophy of sensorimotor activity

Rejecting the static existence of experience is critical to process philosophy. Process philosophy emphasizes the continuity of the immediate past and the immediate future in the actualization of experience, meaning there is no state in experience but rather a way of experiencing. The actualization of a process is the exfoliation of the real by actualizing the Virtuals that in turn are left behind as the process continues (Rescher, 2000, p. 22). As will be emphasized later, the processes that have not been unfolded, i.e. the Virtuals, play a critical role in the architectural experience. SMC already hints that the Virtual supports the actual. Conceiving architectural experience through the lens of process philosophy, at once, the architectural experience can only be captured in its transtemporal dynamics (Djebbara, Fich and Gramann, 2019), and these elements of the transtemporal dynamics entail the body.

Ontologically, process philosophy casts the nature of processes as Janus-faced, i.e. they face inward and outward. In principle, the internal structure of a process may be constituted by smaller processes, while being itself constituted by a greater process. A hierarchical organization of short-term process and long-term processes is central to the ontology of process philosophy (see e.g. Klein, 2014). As will be elaborated in Section 3, the hierarchical structure is indeed embedded in the structure of the brain. For now, architectural experience has been drawn through the combined lenses of phenomenology and SMC emphasizing how the experience of architecture depends on the bodily structure and continuity in sensation and action. SMC, however, is faced with the challenge of providing the neuronal basis on which the mastery of sensorimotor contingencies unfolds. To this end, Section 3 briefly reviews active inference as corresponding to the action-perception loop.

### 3. A process theory of action-perception

Process philosophy casts the sun not as a thing, but as an enduring fire and Heraclitus' river a continuous flow and not a fixed object. Everything the world is made of is a continuous process of activity and change that mutually affect one another. The major point of SMC is that a cortical structure 'alone' cannot account for perception but depends on reciprocal causality with the body, brain, and environment (Clark, 1999; Noë, 2004; Varela, Thompson and Rosch, 2016). Since our perception of the world can only be as good as the information garnered by sensory systems, defective sensory systems can theoretically enact a multiplicity of worlds. Assuming intact sensory systems, the enacted information reveals something about the world—however, it was argued throughout phenomenology and SMC that our perception of the world is not provided by the sum of bottom-up sensory. Instead, there is a context-sensitive top-down prediction based on prior experiences that guides our active 'palpation' of the sensorium.

The current trend in cognitive neuroscience and philosophy considers the brain as an organ of inference, generating (Bayes optimal) predictions (Rao and Ballard, 1999; Friston, 2010; Hohwy, 2013; Clark, 2015). Embodied perception is similarly gaining momentum in the predictive framework (Thompson and Varela, 2001; Seth, 2014; Varela, Thompson and Rosch, 2016; Gallagher and Allen, 2018). Generally, the predictive brain framework draws on the fundamental condition for survival in any biological system, which is the maintenance of physiological balances ensuring the necessary preconditions for survival (Friston, 2005; Damasio, 2010). By maintaining physiological—and other existential—states within bounds, the biological organism resists disorder, dissipation, decay, and ultimately death.

The free energy principle (FEP) is a first principle description of the requisite dynamics, describing how the biological organism resists the tendency to disorder by minimizing the surprise associated with the sensory sampling of the environment (the surprise is variously referred to as self-information, surprisal, prediction error, etc). The following section considers a computational account of action and perception[3] by briefly reviewing a corollary of the free energy principle (Friston, 2010), namely 'active inference' (Friston *et al*, 2017). By computational it is meant algorithmic. Active inference is 'algorithmic' in the sense that it is a realizable biophysical process that performs approximate Bayesian inference (and implicitly surprise minimization), rather than an abstract 'algebraic' formula.

Introducing approximate Bayesian inference eludes the intractable issue of estimating all possible values of hidden states or latent states of the world generating sensory samples. For the embodied brain to infer the cause of sensory samples, it instead performs an approximate Bayesian inference, known as Bayesian belief updating of one form or another (e.g., predictive coding or belief propagation). Applying Bayes formula to infer causes of sensation rests on the following formulation:

$$p(s|o) = \frac{p(o|s)p(s)}{p(o)} \leftrightarrow p(o) = \frac{p(o|s)p(s)}{p(s|o)}$$

Importantly, Bayes' theorem quantifies the probability of a hidden state given an observation. The probability of an observation 'p(o)' functions as a quantity that represents the score of the model. This score is known as Bayesian model evidence or marginal likelihood—and is simply the probability of an outcome under a model of how that outcome was generated. The imperative for survival is to ensure outcomes are not surprising and their probability is maximized. This is the same as maximizing the evidence for the organism's generative model; sometimes referred to as self evidencing (Hohwy, 2016). The embodied brain must therefore sample outcomes that provide evidence for its model of the world; namely, evidence for its own existence. However, estimating the evidence explicitly is usually an intractable issue, as there may exist infinite causes of each sensory signal (Figure 3). This brings us to approximate Bayesian inference as a realizable process that brings Bayesian beliefs about the causes of sensory outcomes as close as possible to the true posterior distribution 'p(s|o)' based upon Bayes rule above. In other words, inference means finding the most likely probability distribution over the states generating outcomes, given those outcomes and a generative model. By introducing a free energy functional of Bayesian belief about states of the world, one can convert an intractable *inference problem* into a tractable *optimization problem* that can, in principle, be solved using neuronal dynamics — cast as a gradient flow on free energy (Friston, 2010).

---

[3] Unnecessary technical details are relegated to footnotes.

In summary, the model evidence 'p(o)' quantifies how well the model predicts observations. It is also known as the marginal likelihood because it is the sum of all possible hidden states given a sensory signal (it marginalizes the hidden state). Although the evidence cannot be estimated directly, there exists a free energy approximation to model evidence that is a function of Bayesian beliefs. By changing Bayesian beliefs—encoded by neuronal activity—to minimize the free energy, the surprise is minimized and free energy becomes the (negative) logarithm of model evidence. At the same time, the Bayesian beliefs become posterior beliefs about their causes of sensory outcomes.

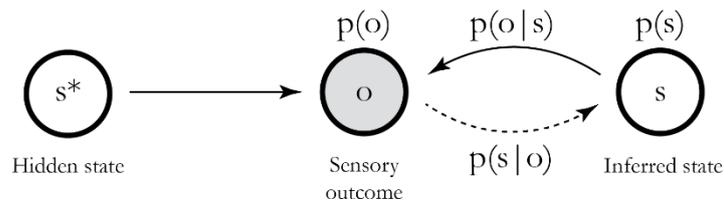

**Figure 3**. Architecture functions as a hidden state that the brain must infer to perceive. The hidden state generates a sensory outcome, which in turn enters a Bayes inferential loop that infers the cause of the sensory outcome resulting in an inferred state about the hidden state. 's*' designates the true state, as opposed to the inferred state, which is 's' without an asterisk. Technically, 's' is a hypothesis about 's*'. The inference is the process of computing p(s|o) – i.e., the probability of each hypothetical state given the observations. While a subtle difference, this highlights that the result of inference is a (posterior) probability distribution, not a single state. However, the states described by this distribution do not necessarily include the hidden state in the outside world.

## 3.1. Two modes of perception

In brief, active inference cast the process of perception as a process of inferring the external causes of sensory signals. When dealing with models of discrete 'states of affairs' it usually considers a hierarchical generative model based on a discrete Markov decision process (MDP). In this setting, perception of the environment is inferred by reducing uncertainty about hidden or latent states in the world under the assumption observable outcomes were generated by this process [4]. Crucially, a Markov decision process puts action into the generative model because the (Markovian) evolution of states over time depends upon choices, decisions, or actions.

This consequently means that a living organism does not have direct access to the world, but must use its sensory organs to infer states of affairs in the environment. Perception thus emerges from the generative model of latent, hidden or external causes that best explain the sensorimotor dynamics. This inference to the best explanation can either involve belief updating to explain current sensory samples to minimize surprise or, alternatively, one can change the sensory signals by acting in the environment to make them less surprising (i.e., bring sensations into line with predictions). These complementary facets of minimizing surprise or free energy can be thought of as perception and action respectively. However, they both work hand-in-hand in the service of active inference or self evidencing. The ensuing relationship between action and inferring the causes of sensations constitutes active perception of the environment, which is well in line with phenomenology and SMC.

---

[4] Indeed, the free energy principle builds primarily on information theory where the uncertainty about the observations is considered as Shannon entropy: $H(x) = -\sum_x p(x) \cdot \log p(x)$. This was initially developed for estimating the surprise of signals. Information in this sense is defined as the 'newness' in the provided knowledge that was not earlier predicted—hence, surprise.

In summary, there are two aspects to perceptual inference or synthesis. One is to approximate the sensory information by inferring the causes of sensory signals. Another is to alter the sensory signals by acting on the environment to solicit the kind of signals that minimize surprise and uncertainty. The aim is to unpack active inference according to phenomenology and SMC and thereby provide a deeper insight into the dynamics of sensation, action, and perception. Admittedly, although SMC provides an account of perception by way of anticipation of sensations if one were to act, it did not, however, provide a corresponding neuronal process theory. Therefore, if the SMC framework can be cast as variational Bayesian inference under a Markov decision process, the following provides a quantitative account of (i) perception by minimizing surprise, and (ii) action which minimizes expected surprise (i.e., uncertainty).

### 3.1.1. The mode of approximation

Sensory information is continuously solicited by sensory organs. The implicit sensory enaction rests on top-down predictions that quantify surprise, often thought of as a prediction error; namely, the mismatch between what was observed and what was predicted. Therefore, perception of the world rests on the ability to minimize prediction errors using the continuous influx of sensory information. In predictive coding formulations of active inference, the error estimation is simply given by the difference between the top-down prediction and the incoming bottom-up sensory signal. More generally, we can cast this kind of evidence accumulation or Bayesian belief updating in terms of Bayesian beliefs about hidden states—denoted 'q'—that approximate the true posterior probability—denoted 'p'. The main goal of perception is to approximate the true probability by minimizing the difference between 'q' and 'p' where 'q' is encoded by neuronal activity.

The quality of this approximation takes the shape of Kullback-Leibler divergence (KL-D), which is a measure of how one probability distribution (or density) differs from another (Kullback and Leibler, 1951)[5]. In other words, the KL-D describes how well one probability distribution matches another, where KL-D equals zero corresponds to a perfect match. KL-D measures the anticipated number of extra bits required to code samples from 'p' when using an approximation 'q'—thus, KL is not a direct measure of distance (Bishop, 2006, p. 55), but rather an indirect measure of uncertainty, reflecting how surprising a set of sensory outcomes are on average. Additionally, due to the nonnegative properties of KL-D[6], it is possible to apply Jensen's inequality to ensure that a minimum exists. Consequently, it is possible to bring 'q' towards the true probability distribution 'p' by minimizing a (free energy) functional that includes this divergence. By subtracting the log-evidence from the divergence, we obtain Feynman's (1972) variational free energy that represents a bound on the log-evidence (see Appendix for more details).

In short, free energy is the divergence between approximate and true posterior beliefs minus log-evidence. Minimizing free energy minimizes the divergence between Bayesian beliefs and the true posterior beliefs while, at the same time, minimizing negative log-evidence. Note that the divergence depends upon Bayesian beliefs (i.e., perception), while the log-evidence depends upon outcomes that are actively solicited (i.e., action).

---

[5] The KL-D is defined as: $D_{KL}(p||q) = H(p,q) - H(p)$, where 'p' is the reference probability being compared to 'q', and $H(p,q)$ is a cross entropy. The outcome reflects how well the probability 'q' approximates the probability 'p'. Relevant to the purpose of active inference, the KL-D can be written in the expectation, discrete and continuous form too: $D_{KL}(p||q) = E_p\left[\log\frac{p(x)}{q(x)}\right] = \sum_x p(x)\log\frac{p(x)}{q(x)}$ if $x$ is a categorical variable, or $E_p\left[\log\frac{p(x)}{q(x)}\right] = \sum_x p(x)\log\frac{p(x)}{q(x)} = \int p(x)\log\frac{p(x)}{q(x)}dx$ if it is continuous.

[6] Negative log is convex where Jensen's rule of inequality applies.

Another useful formulation is in terms of accuracy and complexity:

$$F = \underbrace{D_{KL}(q(s)||p(s))}_{complexity} - \underbrace{E_q[\ln p(o|s)]}_{accuracy}$$

The Bayesian beliefs that minimize free energy therefore provide an accurate account of the sensory outcomes in the simplest way possible. Here, complexity is the divergence between approximate posterior (Bayesian) and prior beliefs; in other words, how much one has to 'change one's mind' to explain the data at hand. This formulation reveals the nature of Bayesian belief updating: active inference or self-evidencing minimizes the variational free energy, which can either be thought of (i) minimizing a KL-D bound on surprise or (ii) providing the simplest and accurate account of observed outcomes.

**The mode of action**

As noted above, action can directly minimize free energy by changing the outcomes that are sampled. However, actively minimizing free energy requires the free energy expected following an action. This is 'expected free energy'. In active inference, actions are predictions themselves, that rest upon Bayesian beliefs about policies—that are composed of action sequences to act. Each policy has an expected free energy, which scores how likely it is that this policy is being pursued. The lower the expected free energy of a policy, the more likely it is to be selected.

In contrast to perception as a process that minimizes surprise, action minimizes expected surprise, namely uncertainty, i.e. expected free energy. Since active inference is based on information theory, the uncertainty corresponds to entropy, where the expected entropy can be given decomposed into extrinsic (pragmatic, goal-seeking) and intrinsic (epistemic, information-seeking) parts (Kaplan and Friston, 2018). Depending on the person in question and the associated prior beliefs, one may have an information-seeking and epistemic behavior that is comparable to the wandering through a museum—foraging for information—or a pragmatic behavior that is comparable to seeking the restaurant. Epistemic behavior seeks towards increasing salience and novelty, while pragmatic behavior seeks towards a particular outcome cast in terms of prior preferences.

Importantly, both behaviors depend on the outcome of the sensorimotor dynamics. Aligning with SMC, perception depends on the incoming sensory and the predicted observations under a particular policy. Therefore, the sensorimotor dynamics can be conceived of underwriting expected free energy. The expected free energy is given by the free energy as associated conditioned on action policies, i.e. the expectation over hidden states and yet to be observed outcomes, i.e. 'q(o, s)'. In the same way that we can split free energy into complexity and accuracy, the expected free energy can be split into *risk* and *ambiguity*. We can also express expected free energy as the expected KL bound on log evidence (intrinsic value) and the expected log evidence per se (extrinsic value):

$$G(\pi, \tau) = \underbrace{D_{KL}(q(o_\tau|\pi)||q(o_\tau))}_{expected\ risk} + \underbrace{E_{\tilde{q}}[H[q(o_\tau|s_\tau)]]}_{expected\ ambiguity}$$

$$= \underbrace{E_{\tilde{q}}[\ln q(s_\tau|o_\tau,\pi) - \ln q(s_\tau|\pi)]}_{intrinsic\ value} - \underbrace{E_{\tilde{q}}[\ln q(o_\tau)]}_{extrinsic\ value}$$

In brief, the expected free energy comprises the expected cost, which is the divergence between the expected outcome under an action policy and the prior preference, and the expected ambiguity[7]. Perhaps the most intuitive way to understand the selection of actions that minimize expected free energy is to think about the way we deploy our visual gaze. This has two components; namely, responding to epistemic affordance in the form of intrinsic value to resolve uncertainty about the visual scene, based upon the salience or intrinsic value of 'looking over there'. At the same time, there will be a pragmatic value in avoiding unlikely sensations, such as looking directly at the sun. A complementary perspective on this is in terms of risk and ambiguity: "I will resolve uncertainty by minimizing ambiguity while avoiding the risk incurred by looking directly at the sun".

Following active inference, one should not expect sensorimotor-related surprises in scenarios that are predictable in practical terms—however, one should instead expect an emerging sensorimotor pattern culminating in a practical experience relative to that perception and that predicted action. The architectural experience is not present; it is predicted.

Interpreting sensorimotor contingencies in light of active inference, perception emerges from the dynamics of minimizing the uncertainty about beliefs under a particular action. According to O'Regan and Noë, the kind of knowledge in SMC is practical, in the sense that the perceiver knows how to interact with the percept; it is a skillful deployment of specialized practical knowledge relative to the sensory organ. It is knowing "[…] the ways in which stimulation in a certain sense modality changes, contingent upon movements or actions of the organism" (Hutto and Myin, 2013, p. 25). Such dynamics encompass a multiplicity of modalities as all sensory modalities are expected to be affected by an action. The enmeshment of sensory into a 'whole' experience is beyond the scope of active inference alone—however, combined with SMC, who provides an account of the 'binding' of sensory modalities, the framework reaches novel heights (Seth, 2014). From this brief review of active inference, it is exhibited that the computational model grants a phenomenological alignment that corresponds to SMC.

### 3.2. A computational neurophenomenology?

The term 'neurophenomenology' emerged in a paper by Varela (1999) when investigating time and perception. It is an attempt at correlating the continuous phenomenal experience with brain activity and thereby operating between quantitative and qualitative methods (Olivares *et al.*, 2015). Meanwhile, computational phenomenology provides an explanatory framework of perceptual experience as the outcome of a functional system that translates physical stimulus to meaningful percepts (Harlan, 1984). According to the current framework, the translation from physical stimuli to meaningful and phenomenal percepts involves the practical coherence between brain, body, and environment. A computational neurophenomenology, therefore, reflects a research domain that seeks to go beyond neurophenomenology by adding a computational aspect, while going beyond computational phenomenology by adding an aspect of neuroscience. Notably, the framework is a process theory that attempts to understand perceptual experience through the dynamics between action and perception.

Both phenomenology and SMC have been introduced as processes theories while active inference has been introduced as a belief updating or propagation scheme can be realized in physical processes (described in terms of securing evidence for generative models of the world). How do we identify and characterize a process? Although active inference can be stated in continuous form, the essential property of the process theory for discrete states (e.g., the Markov decision process generative model above) is its recursive temporal structure. Continuity is provided in phenomenology by the temporal structure of experience as provided by Husserl (2001a). The experience of a specious present depends on the unity of the immediate past and immediate future (Figure 2). This is indeed reflected in the process of active inference because the current state did not occur juxtaposed as "[…] just one darn thing after another" but rather as an interpenetrated causal process that stretches in time as it is actualized (Rescher, 2013, p. 29). Besides being actualized, a process also needs to respect a 'space-less transition' principle. The space that usually defines a transition is just as space-less as the space between continuous numbers when it comes to experience[8]—in fact, the numbers interpenetrate (Bergson, 2001, p. 97). By approaching the issue of experience through non-linear transitions or dynamics, the problem is posed in time rather than in space. Indeed, it is the space-less transition principle that makes the architectural experience a stream of indivisible quality as opposed to the infinitely divisible space. According to Bergson, since architecture involves indivisible successions of movement animating space, it is an 'art of experience', rather than an 'art of object'. For this particular reason, Bergson stated that issues "[…] relating to subject and object, to their distinction and their union, should be put in terms of time rather than space" (Bergson, 2004, p. 77). How is active inference (com-)posed in time? This is exactly the endpoint of scene construction under active inference. In other words, inferring the constructed environment rests upon a sequence of (usually visual) palpations that successively resolve uncertainty about the causes of sensations such that the next epistemic act depends on sensitivity upon the previous action. This speaks to a constructive and temporally extended form of evidence accumulation that can be cast in terms of Bayesian belief updating. The very appeal to 'updating' speaks to the quintessential assimilation in "time rather than space".

---

[7] Note that $\tilde{q}$ is the predicted posterior distribution given a probabilistic mapping (likelihood, $\Lambda$) from hidden states $s_\tau$ to outcomes $o_\tau$ under a policy $\pi$ at the time $\tau$ in the future: $\tilde{q} = p(o_\tau|s_\tau)q(s_\tau|\pi)q(A)$.

Within active inference, belief updating rests upon likelihood mappings between causes and consequences—and, crucially, transitions among hidden states of the world, or may not, be under active control (Figure 4). In active inference under Markov decision processes, this kind of belief updating is usually formalized using generic models: e.g. an 'A' matrix corresponds to the likelihood, a 'B' matrix corresponds to probabilistic transitions determined by a policy encoded by 'π', a 'C' matrix corresponds to the prior preferences about outcomes, a 'D' matrix corresponds to the initial beliefs before any sensory observation, an 'F' function corresponds to the variational free energy, a 'G' function corresponds to the expected free energy and an 'H' matrix corresponds to the entropy of likelihood (Parr and Friston, 2019).

Under this sort of parameterization, the dynamics of perception and action become rather straightforward. Accordingly, the belief update of Bayesian beliefs about hidden states become[9]:

$$q(s_\tau|\pi) = Cat(s_{\pi,\tau})$$

$$s_{\pi,\tau} = \sigma \left( \underbrace{\ln B_{\pi,\tau-1} s_{\pi,\tau-1}}_{prior} + \underbrace{\ln B_{\pi,\tau+1} s_{\pi,\tau+1}}_{predicted} + \underbrace{\ln A \cdot o_\tau}_{likely\ outcome} \right)$$

Here, given an action policy at a point in time, the hidden state is inferred through a softmax function, 'σ', of the sum of the prior moment, a prediction of the upcoming moment, and the likelihood. There is no point in time during active inference where the immediate past and immediate future are not integrated by being simply juxtaposed. Therefore, the discretizing reflected in active inference is only as real as the seconds and minutes of our continuous-time. Beyond the aim of resolving perception through action, the take-home message of this treatment is that active inference operates under the space-less transition principle of process theory making the framework compatible with phenomenological frameworks.

---

[8] See for example Zeno's paradoxes on Achiles and the turtle and the paradox of the flying arrow.
[9] The notation 'Cat' means a categorical distribution, whose sufficient statistics are a vector of probabilities for each alternative state.

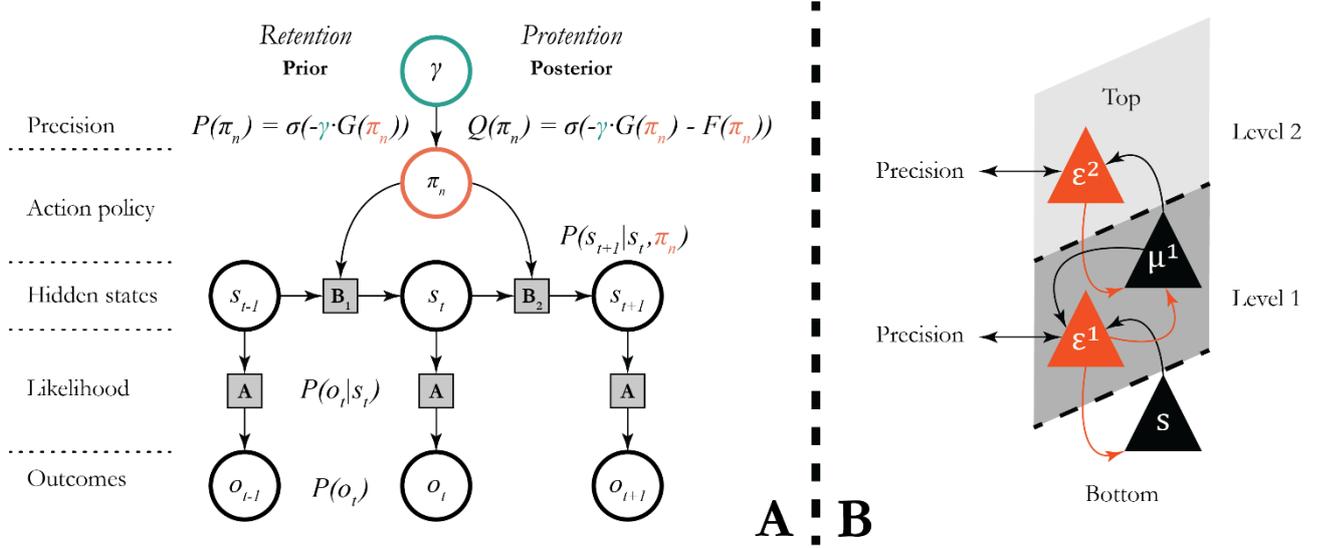

**Figure 4. A. A discrete state-space Markov decision process.** This diagram illustrates a probabilistic generative model. The hidden states (s) generate sensory observations (o) using the likelihood matrix (A). The transition from one state to the next in time depends on the transition matrix (B), which further depends on the action policy. In other words, changes in observations depend on the selected kind of action, which in turn depends on the policy. The probability of an action policy depends on the expected free energy and precision, which is given by the softmax function of the expected free energy. $\gamma$ designates the precision of beliefs about policies, $\pi$. There are many alternative policies ($\pi_n$) considered at the same time, and the precision scores the degree to which their probabilities differ. **B. A continuous state-space predictive coding scheme.** This figure schematically designates 2 hierarchical levels in the brain. The red triangles represent PEs while the black triangle represents expectations. Sensory enters at the bottom while predictions emerge from the top. The $\mu$ represents the expectation, which is adjusted by the $\varepsilon$ that in turn represents the difference between sensory input and the prediction (Friston, 2019). Note that the 's' in **A** and **B** are not equivalent. In **A**, this symbol refers to a hidden state. In **B** it refers to a sensation (loosely analogous to the outcomes in **A**).

## 4. Affordances in architecture

To apply to architecture, the environment is integrated through possible actions. Since actions are restricted by the environment, the structure of the body, and the individual capabilities, they reflect the designed environment with a reference in the body. According to Section 2 and 3, the architectural experience is a transtemporal consequence of the continuous collection of sensory information, where the modalities differ according to their respective sensorimotor contingencies, e.g., when we can look, where we can move, and how we palpate. Additionally, our grip on the environment is reflected in the extent of action policies. Under this proposed computational neurophenomenological framework, the ecological concept 'affordance' corresponds to the expected free energy—that determines the probability that "this is the kind of epistemic or pragmatic policy that I will pursue". In turn, this speaks to distinct different kinds of affordance; namely, epistemic and pragmatic affordance. As elaborated in Section 3, the epistemic behavior seeks to increase novelty, while the pragmatic behavior aims to bring about preferred observations. Such 'exploration' and 'exploitation' refer to how action policies are evaluated (Kaplan and Friston, 2018). This distinction is useful when considering sensorimotor and ideomotor dynamics (Hommel, 2009).

### 4.1. Sensorimotor and ideomotor views

In psychological terms, the ideomotor views reflect motor actions that are intended (idea-guided), whereas sensorimotor views reflect actions suggested by the environment (perception-guided). Conservative sensorimotor views conceive action as re-actions to stimuli—so that behavior is dominated by environmental cues. Physical causes are favored above mental causes. In contrast, ideomotor views emphasize goal-directed actions that originate from autonomous beliefs about action. The goal is central in such a view. Indeed, these concepts provide a helpful distinction for discussion. During an architectural experience, actions are dependent on both external and internal factors because actions are structures that link movement to goals, and goals to movements (Prinz, 1997). Behavior necessarily emerges from the dynamics of these views. For instance, in an enclosed square space, the behavior of a human being cannot be accounted for by the sensorimotor affordances of the built environment alone. It is, however, safe to say that any behavior will be with respect to the affordances of the built environment. Behaviors such as wandering around, doing push-ups, jumping jacks, etc., these are ideomotor actions that are highly history-dependent, i.e. the centered around the history of the individual who has encultured certain actions under certain situations. Behavior can thus be accounted for by considering the combination of external Virtual actions and internal affordances. This is precisely where the Bayesian brain under active inference operationalizes SMC[10] (Figure 5).

Although recent advances in neuronal process theories have led to different algorithmic models of the Bayesian brain, the general approach is based on a hierarchically organized generative model, where the hierarchical layers couple through top-down (TD) and bottom-up (BU) signals. In predictive coding schemes, hierarchical levels communicate bidirectionally and encode predictions and prediction-errors that are further weighted by their precision. When applied to perception and action, the purpose of the generative model is to infer by TD predictions the external causes of the incoming BU sensory signals by minimizing free energy. To recapitulate, the free energy refers to an information theory measure of surprise, i.e. the amount of prediction-error (known formally as self-information). Thus, the structure of the inference is reminiscent of sensorimotor and ideomotor dynamics. The grip on the environment corresponds to the state space of action policies. Action policies reflect the actions that compete in the current situation, biased by both sensorimotor and ideomotor processes, to minimize expected free energy. In unknown spaces, action policies under consideration are scanty and with high uncertainty, as opposed to known spaces, with a plethora of action policies and with high certainty. As exploration unfolds, the precision of beliefs about action policies increases and thereby improves the grip on the environment. Paraphrasing Aristotle, the things we have to learn before we can do them, we learn by doing them. Affordances depend on the sensorimotor dynamics, while a grip on the context where actions can unfold depends on actions. Interestingly, sensorimotor dynamics concerning architectural affordances have been investigated on a neuronal level.

---

[10] An alternative to the contingency-law could be provided by the 'binding by synchrony' hypothesis so that the mastery can be defined as the practical knowledge in how to map cross-modality information onto each other. Similar to SMC, the main argument is that the perceptual experience evident in architectural experience is a matter of temporal coherence rather than states of neuronal assemblies (Buzsáki, 2006, pp. 250–261). However, reviewing similarities and differences between SMC and "binding by synchrony" is beyond the scope of this paper.

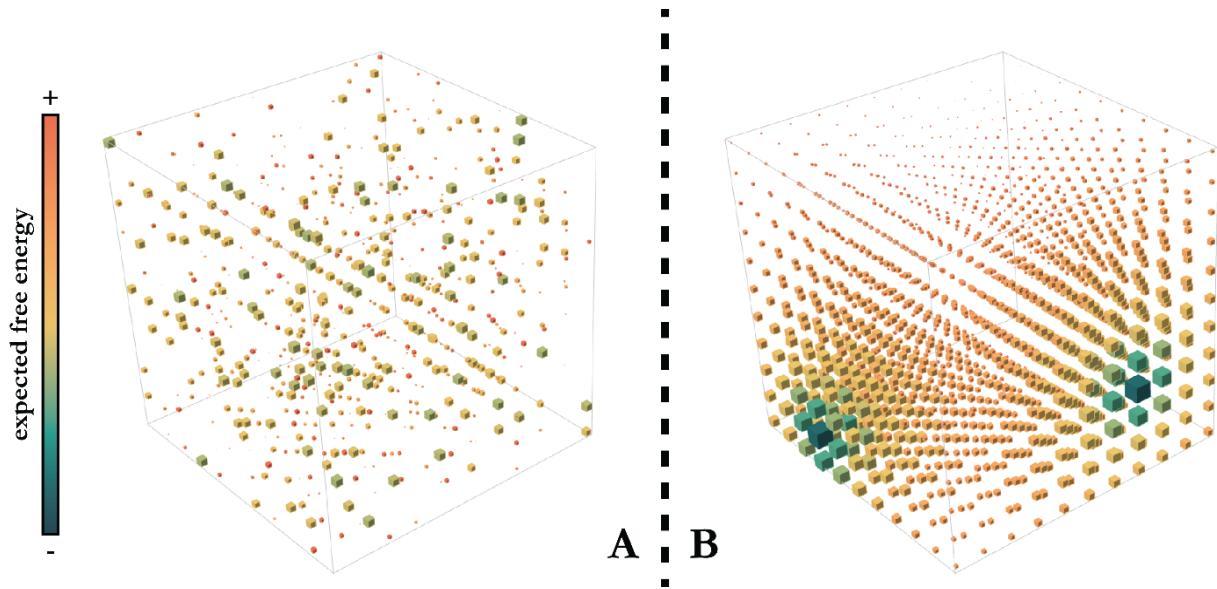

**Figure 5.** State-space in three dimensions of action policies. Each policy is designated by a box where the size and color are relative to the expected free energy. **A.** A fictive example of high expected free energy among action policies without any apparent attractor. Such a state suggests exploring to reduce uncertainty. **B.** A fictive example of two attractors, i.e. competing action policies, in terms of their expected free energy. This could for instance be an ambiguous figure.

## 4.2. Some empirical evidence

Process philosophy is a pragmatist philosophy. It renders the temporal development important to understand the whole. Applying this approach to cognition and scientific progress, it is found that applicative praxis is the best available procedure for assessing scientific progress, i.e. praxis is the arbiter of theory (Rescher, 2000, p. 89). Unfortunately, the use of architectural spaces to understand sensorimotor dynamics in the brain is rather limited.

One way architecture may affect an organism physiologically has been put forward by Fich et al. (2014). Their study entailed inducing psychosocial stress using the Trier-Social-Stress-Test in two different architectural settings that varied in their 'openness', i.e. one space has big windows affording to escape, while the other was fully enclosed. Despite being exposed to the same procedure of psychosocial stress, which was unrelated to the environment, their results show significant and systematic differences in cortisol levels between the two spaces. Fich and colleagues concluded that the enclosed space increased the release of the cortisol hormone due to its restriction on the fight/flight mechanism. By measuring cortisol, which is an indirect measure of hypothalamic activity, their results indicate that sensorimotor affordances are attuned to unconsciously.

Paraphrasing, and in agreement with, Dobzhansky "the brain only makes sense to investigate in light of behavior", active inference suggests that selective attention optimizes (higher confidence of) the expected prediction by modulating the synaptic gain of prediction-errors in predictive coding, i.e. neurons coding for behaviourally relevant prediction-errors should be more responsive. See also (Parr and Friston, 2017, 2018) for the equivalent role of precision in discrete formulations of active inference. This has recently been demonstrated by Djebbara et al. (2019) in a Mobile Brain/Body Imaging experiment using Virtual Reality to change the forms of a transition causing the affordances to change. An electroencephalogram (EEG) was used to analyze visuomotor brain activity. Participants were instructed to transition from one space to another, where the opening between spaces varied offering three different door-sizes. Of the three types of doors, one was too narrow to pass, another was difficult to pass, while the last was easily passable. The results indicate that early sensory brain activity differed as a function of affordances reflecting the lower level of the hierarchical structure in PEs stemming from TD predictions and BU sensory signals. These results suggest that sensorimotor affordances influence perception (Djebbara, 2020). Analyzing the frequency-domain of the same datasets, Djebbara and colleagues were further able to localize alpha desynchronization in the parieto- and temporo-occipital regions, which may serve as markers for sensorimotor integration during interaction with the environment. They conclude that moving in space is to continuously construct a prediction of a world that we perceive as dependent on our action potentials.

Vecchiato and colleagues (2015) also used Virtual Reality and EEG to investigate the sensorimotor responses while perceiving different interior designs. The topographical maps of the EEG data indicated activity in sensorimotor brain areas when participants were experiencing pleasant, novel, and comfortable interiors. This was followed up by an extensive enactive approach to architectural experience suggesting that architectural affordances embody thought, while thought can be structured by architecture (Jelić *et al.*, 2016).

Due to technological restrictions, the few studies are suggestive than conclusive—however, it is crucial that the outcomes of the studies do not dwindle in the theoretical literature but instead engender further empirical testing. Within process philosophy, the validity of a theory depends on how, and how hard, we question nature (Rescher, 2000, p. 83). The progress of this field rests on critical theoretical advances, technological developments, and empirical experimentation. Nature will not reveal more than what we forcibly extract using a particular instrument at our disposal. In doing so, the empirical results are perceived through the lens of a specific theory relative to the limitations of the measuring instrument. Therefore, given the current theoretical proposal, the destiny of a computational neurophenomenology rests on technological developments and empirical experimentation.

## 5. Conclusion: a principle of anticipation

Through a philosophical discussion, phenomenology and SMC were coupled to the experience of architecture, emphasizing the dependency on the bodily structure and continuity in sensation and action. Hereafter, active inference operationalized the laws considered in SMC allowing us to understand the limitations of such contingencies and the underlying neuronal activity bound thereto. Indeed, it is the predictive framework of the brain depicted by (variational) Bayesian inference that is doing the heavy lifting in explaining the architectural experience. It has been shown that a synthesis of phenomenology, SMC, and active inference enables a new perspective on architectural experiences.

Sensory impressions alone (BU signals) were shown to be insufficient in explaining the architectural experience as perception requires a priori (TD) predictions based on earlier experiences from all sensory impressions. It is impossible to perceive without a sense of functionality (affordances) stemming from other sensory modalities. Although the sensory information is visual, the perception relates to all sensory organs, discovering functionalities well beyond visual purposes. The relations between form and perception in architectural experience are enriched by the proposed framework as spatial (architectural) Gestalts that can now be addressed through empirical studies of architecture relative to the human body and sensory. In line with the proposed framework, as the architectural experience is modulated by the affordances it can be deduced that perceptual conjurations can be realized through the manipulation of possible actions. Just like Gestalt psychology uncovered the role of visual sensory in perceptual experience through two-dimensional illusory studies, so we propose that architecture can uncover the role of all sensory in architectural experience.

Scale, distance, luminance, thermal comfort, and other design parameters can all be biased by the affordances of the space. Judging that one space is bigger than another, without this necessarily being true, rests on the affordances of all sensory impressions. For instance, the distance from a house in the Parisian suburbs towards the Eiffel Tower was judged shorter than vice versa (Sadalla, Burroughs and Staplin, 1980; Tversky, 1981). Under the proposed framework, it is suggested that the judgment of distance depends on the 'whole' experience. Approaching the Eiffel Tower is arguably a perceptually more exciting experience than the drive away from it. Further, after a long walk around the Eiffel Tower and vibrant stimulations from the atmosphere of central Paris, one is arguably exhausted on the way back. In this sense, the judgment of distance is an embodied measure through perceptual experience that depends on affordances and anticipation.

The framework proposes a principle of anticipation in architectural experience. In locating the beginning of the architectural experience, one must acknowledge the anticipative nature of each sensory modality. As one proceeds, the continuous collection of sensory information, which depends on the accessibility of the space, contributes to the embodied predictions on the fly. Once arriving at the end of an exhibition in a museum, the final space is indeed modulated by all preceding spaces. Furthermore, the following space hereafter is also modulated by the final space of the exhibition, and so on ad infinitum. The experience of space thus depends on its coherence with its contextual space. Moving in space is to continuously construct a prediction of a world that we perceive as dependent on affordances. These are in turn reflected in cortical oscillations. By designing our environments, architects design cortical activity. There is still a long road ahead to adequately understand how affordances and the dynamics of architectural form impact the multiple levels of the brain.

### 6. Future research

Architectural experience has frequently been linked with neuroaesthetics (Vartanian *et al.*, 2013; Skov *et al.*, 2017). Hedonic psychology and affective neuroscience pursue to understand the underlying mechanisms of pleasure, and further how it is represented in the brain (Berridge and Kringelbach, 2011). Hedonic value is central to the field of neuroaesthetics and can be integrated with the proposed framework. One way for such integration is to allow the hedonic value to bias the expected free energy when exploring space. For instance, exploring an exhibition at a museum, the sight of a specific painting from afar that holds a hedonic value may attract the exploration in that direction. Within active inference, prior beliefs (e.g., the C matrix) present themselves as a candidate for integrating hedonic value as they subsume prior preferences about outcomes. Another way for such integration is to determine the hedonic value of the actions themselves. For instance, recent neuroscientific research in dance indicates a therapeutic effect to combat neurodegeneration (Barnstaple *et al.*, 2020). In this sense, there may be an aesthetic value in the actions themselves that can be evaluated. Although these areas remain complex to address, the proposed framework allows conceiving aesthetic appraisal as a form of sensory exploration where the extent of action policies and minimizing entropy are suggested to play a central role.